\documentclass[aps,prl,twocolumn,superscriptaddress,amsmath,amssymb]{revtex4-1}
\usepackage{graphicx}

\begin{document}

\title{Searching for Non-Abelian Phases in the Bose-Einstein Condensate of Dysprosium}
\author{Biao Lian}
\affiliation{Institute for Advanced Study, Tsinghua University, Beijing, 100084, China}
\author{Tin-Lun Ho}
\affiliation{Department of Physics, The Ohio-State University, Columbus, Ohio, 43210, USA}
\author{Hui Zhai}
\affiliation{Institute for Advanced Study, Tsinghua University, Beijing, 100084, China}
\date{\today}

\begin{abstract}
The recently realized Bose-Einstein condensate of Dysprosium will become a spin-$8$ spinor condensate at ultralow magnetic fields. In such a high spin condensate, many phases with different symmetries can exist. Among them the most interesting ones are those with non-abelian point group symmetry. In this letter we discuss the variety of symmetry phases in a spin-$8$ condensate resulting from numerical solutions of the Hamiltonian. We show that these symmetries can be determined uniquely from the measurements of density population on each spin component in an ultralow magnetic field, together with the measurements of the collective modes in the zero-field limit. This method can also be applied to Bose-Einstein condensate of other magnetic atoms, such as Cr and Er.
\end{abstract}
\maketitle

Recently Benjamin Lev's group has successfully realized Bose-Einstein condensate of Dysprosium (Dy)\cite{Dy}. Dy is a complex lanthanide atom with totally $10$ $f$-shell electrons.  It has electron spin $S=2$, orbital angular momentum $L=6$, and nuclear spin $I=0$. The total spin ${\bf F}={\bf L}+{\bf S}$ becomes $F=8$ due to the Hund's rule. The spin of Dy can be easily polarized by an external magnetic field, which will then give rise to
 strong magnetic dipolar interactions. However, it is technically possible to reduce the magnetic fields to very low values so that the system is essentially depolarized, as recently  achieved in the Cr condensate \cite{Cr}.  In this limit, the spin Hamiltonian has rotational  and gauge symmetry,
 $G=SO(3)\times U(1)$. Bose condensation breaks this symmetry.  However, the resulting state
 can have the symmetry of a  subgroup of $G$, which can be non-abelian for large spin bosons\cite{non-abelian}.
 Condensates with non-abelian symmetry group are particularly interesting, as they have non-abelian topological defects  with  novel physical properties \cite{Mermin}.

At zero-field limit, the actual spin ground state of Dy is determined by nine scattering lengths that are not yet known. It will be overcomplicated to construct a phase diagram with all these unknown parameters. In this letter, we focus on the question of non-abelian phases of Dy condensate, and the simplest scheme to detect them.
While the non-abelian nature of the order parameter is fully manifested in the resulting structure after its line defects cross each other\cite{Mermin}, such experiments are too involved in a cold atom setup at present. In this work,  we propose a much simpler method to determine the symmetry of a spinor condensate.
 It combines the measurements of spin populations and the Bogoliubov spectrum. The former can be achieved with Stern-Gerlach technique while the later can be measured by Bragg spectroscopy. Both techniques are widely used in cold atom experiments today.

{\it Model:} Let $\hat{\psi}_m$ be the bosonic operator for each spin component $m$, $(m=-F,\dots,F)$, the Hamiltonian for  Dy $(F=8)$ is  $\hat{H}=\hat{H}_0+\hat{H}_{\text{int}}$ \cite{FM,Machida},
\begin{align}
\hat{H}_0=&\int d^3{\bf r} \sum\limits_{m=-F}^{F}\hat{\psi}^\dag_m({\bf r})\left(-\frac{\hbar^2\nabla^2}{2M}-B\mu m\right)\hat{\psi}_m({\bf r}) \\
\hat{H}_{\text{int}}&=\int d^3{\bf r}  \sum\limits_{j=0,2,4,\cdots}^{2F}g_{j}\sum\limits_{m=-j}^{j}\hat{A}^\dag_{jm}({\bf r})\hat{A}_{jm}({\bf r})  \label{Hamiltonian}
\end{align}
where $g_{j}=2\pi\hbar^2 a_{j}/M$, $a_{j}$ is the $s$-wave scattering length between a pair of bosons with total spin $j$,
described by the local pair operator  $\hat{A}_{jm}({\bf r})=\sum_{m_1}\langle j,m|F,m_1,F,m-m_1\rangle\psi_{m_1}({\bf r})\psi_{m-m_1}({\bf r})$. Bose statistics, however, implies $j=0,2, ..16$ only.
$B$ is a tiny magnetic field along $\hat{z}$ direction, and the magnetic moment $\mu=5\mu_{\text{B}}/4$ for Dy \cite{supple}.

The energy of a condensate  $\langle\psi_{m,{\bf k}=0}\rangle=\sqrt{n_0}\varphi_m$ is
\begin{equation}
\mathcal{E}=\sum\limits_{j=0,2,\cdots}^{2F}g_{j}n^2_0\sum\limits_{m}|\mathcal{A}_{jm}|^2-n_0B\mu \sum\limits_{m=-F}^{F}m|\varphi_m|^2
\end{equation}
where $\mathcal{A}_{jm}=\sum_{m_1}\langle j,m|F,m_1,F,m-m_1\rangle\varphi_{m_1}\varphi_{m-m_1}$, while $n_0$ represents the total density. Using the imaginary time evolution method, we have conducted  extensive searches of possible ground states in zero and small fields for vast ranges of  $g_{j} (j=0,2,\dots, 2F)$. We have further studied the Bogoliubov spectra of the ground states to ensure their stability, which also turns out to be very useful for determining the symmetry of the state.

{\it Majorana Representation:} A very useful way to describe the condensate wavefunction $\{ \varphi_{m}\}$ is to use the Majorana representation of spin states \cite{Majorana}. Recently, many authors have applied this method to study Bose condensate with high spins \cite{non-abelian,highspins,Schwinger}. This representation is most conveniently described in terms of Schwinger bosons\cite{Schwinger}. This method presents the spin operators in terms of bosons  $\hat{a}$ and $\hat{b}$, such that
 $\hat{F}_x=(\hat{a}^\dag\hat{b}+\hat{b}^\dag\hat{a})/2$, $\hat{F}_y=i(\hat{a}^\dag\hat{b}-\hat{b}^\dag\hat{a})/2$ and $\hat{F}_z=(\hat{a}^\dag\hat{a}-\hat{b}^\dag\hat{b})/2$. A general normalized spin state can be written as
 $|\varphi\rangle = \sum_{m=-F}^{F} \varphi_{m}|F, m\rangle$, where $|F, m\rangle=  [(F+m)!(F-m)!]^{-1/2}
 \hat{a}^{\dagger F+m}\hat{b}^{\dagger F-m}|0\rangle$. Since the sum in $|\varphi\rangle$ is a homogenous polynomial of $\hat{a}^{\dagger}$
 and $\hat{b}^{\dagger}$ of degree $2F$, it can be factorized as
 \begin{equation}
|\varphi\rangle = \sum_{m=-F}^{F} \varphi_{m} |F, m\rangle =\frac{1}{\mathcal{N}}\prod\limits_{i=1}^{2F}(u_i \hat{a}^\dag+ v_i\hat{b}^\dag)|0\rangle ,
\end{equation}
where $\mathcal{N}$ is a normalization factor to ensure all $|u_i|^2+|v_i|^2=1$.
Absorbing the overall phase factor of the product into $\mathcal{N}$, we can write $u_i=\cos(\theta_i/2)$ and $v_i=e^{i\phi_i}\sin(\theta_i/2)$, so that the spinor $(u_i,v_i)$ can be  represented as a point on a unit sphere denoted by a unit vector $\hat{\bf n}_{i}$ with polar angle $(\theta_i,\phi_i)$.
It is straightforward to show that (i) under a spin rotation, $\hat{\bf n}_{i}$ rotates as a Cartestian vector, and
 (ii) under time reversal,  $(u_{i}, v_{i})\rightarrow (v_{i}^{\ast}, -u_{i}^{\ast})$, which implies
 $\hat{\bf n}_{i}= -\hat{\bf n}_{i}$.  The $2F+1$ complex numbers $\{ \varphi_{m}\}$ (with totally $4F+1$ real variables because of normalization) is now replaced by $2F$ unit vectors $\hat{\bf n}_{i}$ plus an overall phase factor.

\begin{figure}[tbp]
\includegraphics[height=4.3 in, width=3.2in]
{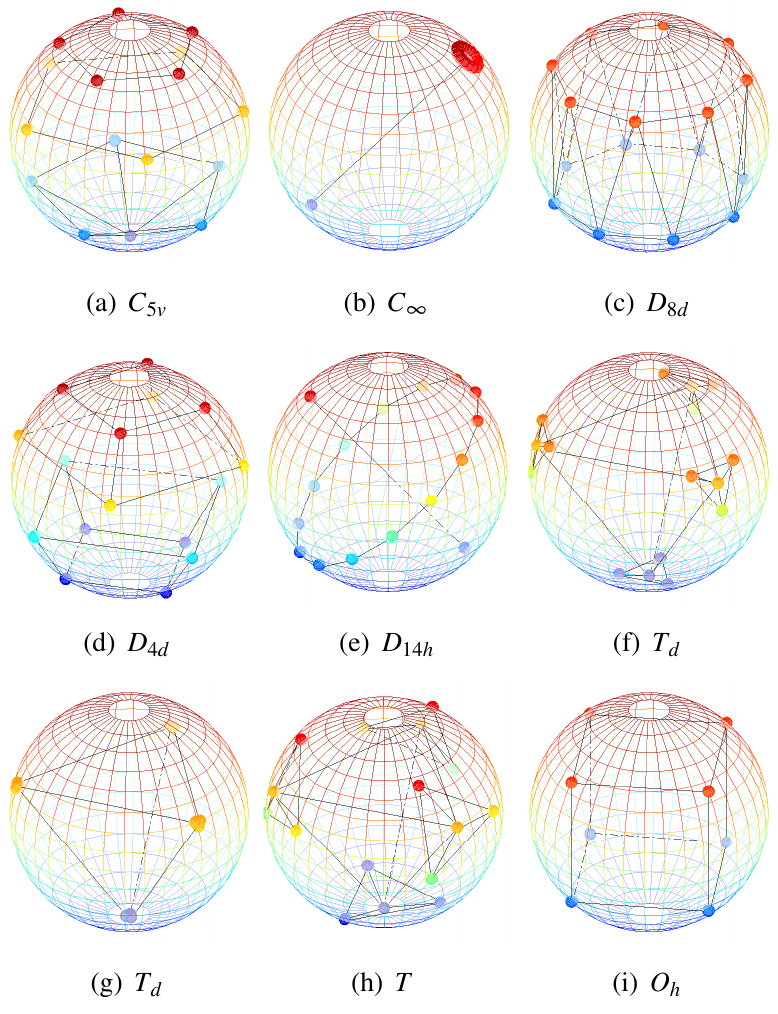} \caption{Majorana representations of spin-8 states with different symmetries.  Their descriptions are given in the discussion of {\it Symmetry Allowed States}. The point with polar angle $(\theta_i, \phi_i)$
on the unit sphere represents the spinor $(u_{i}, v_{i})=({\rm cos}\theta_{i}/2, e^{i\phi_{i}}{\rm sin}\theta_{i}/2)$.
 \label{states}}
\end{figure}

{\it Symmetry Allowed States:}  If $U$ is a rotational symmetry operation, it means the state returns to itself up to an overall phase after being operated on by $U$,  i.e.  $\varphi_{m}\rightarrow \sum_{m'}U_{mm'} \varphi_{m'}= e^{i\alpha}\varphi_{m}$. Since the symmetry group $G$ of $\hat{H}$ is
$SO(3)\times U(1)$, the symmetry of the ground state will be described by one of the normal point groups $E$, $C_n$, $D_n$, $T$, $O$ or $Y$\cite{Mermin}.  $E$ is the identity, corresponding to states with no particular symmetry.  $C_{n}$ denotes an $n$-fold rotational symmetry along certain axis. $D_n$ contains one $n$-fold $C_{n}$ axis plus $n$ two-fold $C_2$ axes perpendicular to  $C_{n}$. $T$, $O$, and $Y$ denote tetrahedral, octahedral, and isocahedral symmetry respectively.  The following properties are useful for our later discussions.

\noindent {\bf (I)} If the vectors $\{\hat{\bf n}_{i}\}$ comes in $({\bf n}, -{\bf n})$ pairs, the state is invariant under time reversal.

\noindent {\bf (II)} A ferromagnetic state (by which we mean $\langle \hat{\mathbf{F}}\rangle\neq0$) breaks time reversal symmetry and can have only one $C_{n}$ axis.
States with more than one $C_{n}$ axis such as those with  $D_{n}$, $T$, $O$, and $Y$ symmetry can not be ferromagnetic, even though they may break time reversal symmetry.

Now let us analyze all the symmetry states:

 (1) $E$: These states have no special symmetry. They are not found as ground states in our energy minimization.

 (2) $C_n$: Dy has $F=8$, hence 16 points $\hat{\bf n}_{i}$ on the unit sphere. So  we  either have  $n\leqslant 16$ or $n=\infty$. The $C_n$ groups can be further classified into $C_{nv}$, $C_{nh}$ and $S_{2n}$\cite{definition}.
 We have found several examples of $C_{nv}$ symmetry. For instance, a $C_{5v}$ state is shown in Fig. \ref{states}(a), in which all $16$ points form three pentagons and one point sitting at the south pole. However, examples with only $C_{nh}$ or $S_{2n}$ symmetry are not found \cite{guess}. We also get states with $C_{\infty}$ symmetry, with $q$
 points $(q<8)$ collapsing into one and other $16-q$ points into the antipodal point. The case of $C_{\infty}$ symmetry with $q=1$ is shown in Fig. \ref{states}(b).
  All the states with $C_{nv}$ and $C_{\infty}$ symmetry break time reversal symmetry, and have ferromagnetic order.

 (3) $D_{n}$:  In this case, either $n=\infty$ or $n$ is a finite integer such that $n\leqslant 16$ for even $n$, and
 $n\leqslant 7$ for odd $n$ \cite{reason}.  $D_n$ symmetry group can be further classified into $D_{nd}$ and $D_{nh}$\cite{definition}. In Fig. \ref{states}(c) and (d), we show two examples of $D_{nd}$. $D_{8d}$ consists of two octagons,  one in the north hemisphere and the other in the south hemisphere. $D_{4d}$ consists of four squares,  two in the north hemisphere and the other two in the south hemisphere.  In Fig. \ref{states}(e), we show an example of $D_{nh}$. For $D_{14h}$, $14$ points are evenly distributed on the equator and the other two are placed on the north and south poles. While some of these states break time reversal symmetry, such as $D_{nd}$, none of the states with $D_{n}$ symmetry are ferromagnetic.
We, however, do not find the ``polar" state with 8 points collapse into the north and the other 8 points to the south pole. We suspect it either does not exist or only occupies a very limited region in the phase diagram.

(4) $T$: We have found one phase with $T$-symmetry and another with $T_d$ symmetry\cite{definition}. They
 occupy a large portion of the phase diagram. An example of the state with $T_d$ symmetry is shown in
 Fig. \ref{states}(f).  Four points are distributed at the vertices of a tetrahedron.  Around each tetrahedral vertex, another three points are distributed at equal distance $d$ from it forming a small regular triangle.  The three vertices of this triangle lie in the three mirror planes passing through this tetrahedral vertex. The distance $d$ varies with interaction and can shrink to zero, as shown in Fig. \ref{states} (g).  We have also found a phase with $T$ symmetry where the vertices of the triangles move away from the mirror plane, as shown in Fig. \ref{states}(h). This transition from $T_d$ to $T$ is second order. Both phases $T$ and $T_d$ break time reversal symmetry.

(5) $O$: This state is described by a cube with two points at each conner, as shown in Fig. \ref{states}(i). This state does not break time reversal symmetry, and is therefore nonmagnetic.

(6) $Y$: This is impossible for Dy, as one can not distribute $16$ points on a sphere with icosahedral symmetry.
On the other hand, such a phase is possible for Erbium (Er) which is $F=6$ and is represented by $12$ points on the sphere\cite{Er}.

All the states with $D_n$ ($n$ finite),  $T$ or $O$ symmetry have non-abelian defects\cite{Mermin}. Hereafter we will discuss how to detect them experimentally.

{\it Detecting the Symmetry:} We propose two measurements ${\bf (A)}$ and ${\bf (B)}$ for probe the symmetry of the system.
Measurement ({\textbf A}) is to pin a $C_{n}$ axis by an ultralow magnetic field along $\hat{z}$. We find numerically that such pinning always happens\cite{supple}. The spin wavefunction along $\hat{\bf z}$ must then be of the form
\begin{equation}
\varphi=(\cdots,0,\alpha,\underbrace{0,\cdots,0}_{n-1},\beta,\underbrace{0,\cdots,0}_{n-1},\gamma,0,\cdots)^T \label{function} \end{equation}
for any finite $n$,
i.e. any non-vanishing component must be separated from the next non-vanishing one by $n-1$ zeros. If $n=\infty$, there will be only one non-zero component. The positions and the magnitudes of the non-vanishing components are also different for different states. This
follows from the fact that  a spin rotation along $\hat{z}$ axes by $2\pi/n$ will change $\varphi$ to  $e^{i\theta}\varphi$, hence  $e^{-i2m\pi/n}\varphi_m=e^{i\theta}\varphi_m$, which can only be satisfied  if $\varphi$ takes the form of Eq. (\ref{function}).

\begin{figure}[tbp]
\includegraphics[height=2.6in, width=2.8in]
{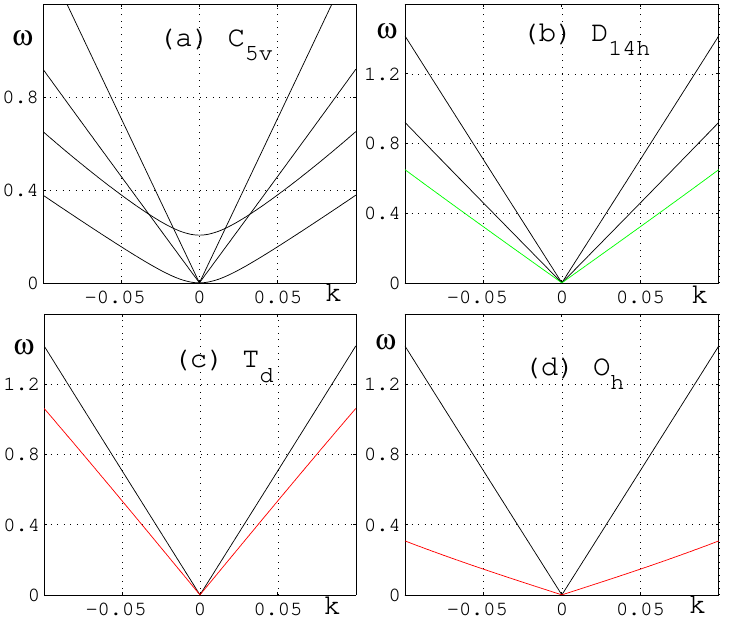} \caption{Low-energy spectra of Bose condensates with different symmetries. $k$ is in unit of $k_0=\sqrt{2m\mu}/\hbar$ and $\omega$ is in unit of $0.1\mu$ ($\mu$: chemical potential). The green (red) line stands for two (three) degenerate modes.
See also Tab. \ref{B}. \label{Bogoliubov}}
\end{figure}

Measurement ({\textbf B}) concerns the low-energy collective modes in zero-field limit. As explained later, for nonmagnetic states with discrete point group symmetry, there will be totally four gapless Goldstone modes with linear dispersion. One corresponds to the broken global $U(1)$ symmetry,
while the other three correspond to broken spin rotation symmetry. If the remaining symmetry contains only one $C_{n>2}$ axis, two spin rotational modes will be degenerate (i.e. only three different dispersions). If there are more than one $C_{n>2}$ axis, all three rotational modes will be  degenerate (i.e. two different dispersions). For ferromagnetic ground states, there will be a quadratic Goldstone mode, as already known from previous study of magnetism\cite{FM}.  While these results are found from explicit calculations of the Bogoliubov modes, they in fact result from the following general considerations.

In the low-energy limit ${\bf k}\rightarrow 0$, the three spin Goldstone modes correspond to simultaneous rotation of all vectors $\hat{\bf n}_{i}$ along  axes $\hat{x}$, $\hat{y}$ and $\hat{z}$. If there is one $C_{n>2}$ axis, say, along $\hat{z}$, such a rotation transforms the fluctuations along $\hat{x}$ and $\hat{y}$ to those along $\hat{x}^\prime$ and $\hat{y}^\prime$, respectively.
Due to the $C_{n}$ symmetry, the fluctuations along $\hat{x}$ ($\hat{y}$) will behave the same as those along $\hat{x}^\prime$ ($\hat{y}^\prime$). On the other hand, the fluctuations along $\hat{x}^\prime$ or $\hat{y}^\prime$ are  linear combinations of those along $\hat{x}$ and $\hat{y}$. It then implies these two modes along $\hat{x}$ and $\hat{y}$ must be degenerate. Furthermore, if there is another $C_{n>2}$ axis, say, along $\hat{x}$ direction, it will imply that fluctuations along $\hat{y}$ and $\hat{z}$ also have the same dispersion. Then, all three rotational modes are degenerate. These results can be applied to different point groups as summarized in Tab. \ref{B}.

\begin{table}[b]
\caption{\label{B} Illustration of collective modes of different symmetry states.  ``$L$" and ``$Q$" stand for modes linear and quadratic in $k$ respectively. ``$2(3)$ dg." is short for two (three) degenerate modes.}
\begin{ruledtabular}
\begin{tabular}{c|c|c|c}
Symmetry&Phase Mode & Spin Modes & Distinct Dispersions \\
\colrule
$E$&$1L$&$1L+1Q$&$2L+1Q$\\
$C_n$&$1L$&$1L+1Q$&$2L+1Q$\\
$D_n$&$1L$&$3L$ ($2$ dg.)&$3L$\\
$T$,$O$&$1L$&$3L$ ($3$ dg.)&$2L$\\
$D_{\infty}$&$1L$&$2L$ ($2$ dg.)&$2L$\\
\end{tabular}
\end{ruledtabular}
\end{table}

{\it Experimental Scheme for Differentiating Different States:}
Applying the Methods $({\bf A})$ and  $({\bf B})$, we can make the following distinctions.

\begin{figure}[tbp]
\includegraphics[height=1.7in, width=3.4in]
{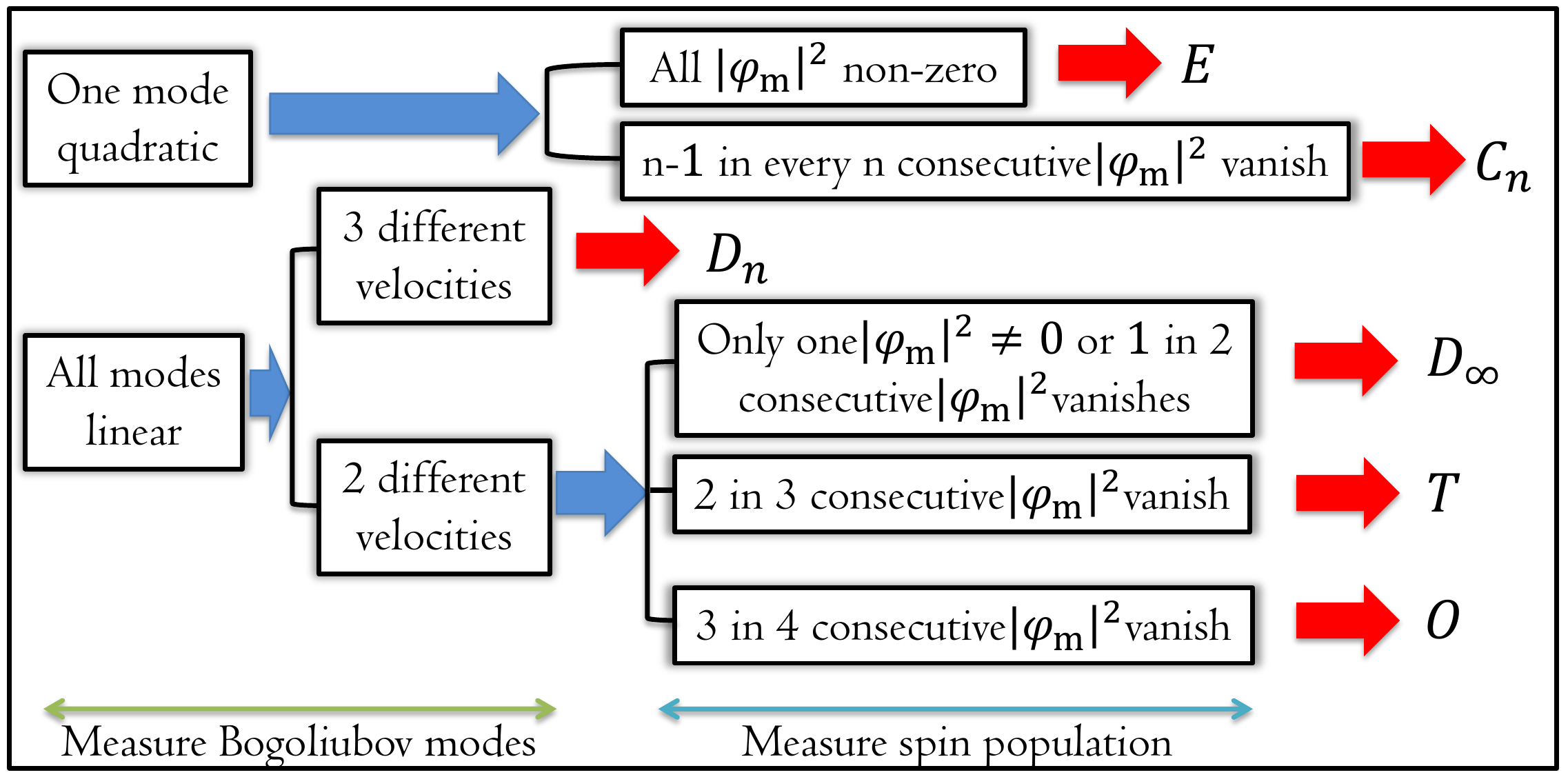} \caption{Schemes for determining the symmetry of different phases. Blue and red arrows denote the measurement of collective modes and spin populations $|\varphi_{m}|^2$ respectively. \label{proposal}}
\end{figure}

(i) Distinguishing $E$ and $C_{n}$ from others: As seen in Tab. \ref{B}. These are the only two ferromagnetic states, and thus have quadratic Goldstone modes. The case of $C_{5v}$ is show in  Fig. \ref{Bogoliubov}(a). Thus, Measurement ${\bf (B})$ can distinguish them from other states. Moreover, $E$ and $C_{n}$  can be distinguished and determined by Measurement ${\bf (A)}$, as $C_{n}$ has the characteristic order parameter as shown in Eq.(\ref{function}).  (Note that our calculations show that an external magnetic field always pin the $C_{n}$ axis of $C_n$ states).

(ii) Distinguishing $D_n$ from $D_\infty$, $T$ and $O$: As seen in Tab. \ref{B}, $D_{n}$ differs from $T$ and $O$ from the number of mode velocities. (See also  Fig. \ref{Bogoliubov}(b)-(d)).
In measurement {\bf (A)}, the untralow magnetic field pins along $\hat{z}$ either $C_n$ or $C_2$ axis of the $D_n$ symmetry, depending on the states \cite{supple}. If $C_2$ axis is pinned, then we can add a small magnetic field gradient ${\bf B}=B[\hat{z}+G_0 (x\hat{x}-z\hat{z})]$. This is effectively equivalent to adding an energy $\epsilon \langle F^2_y\rangle$, with $\epsilon=\hbar^2 G^2_0/(2M)$ \cite{Ho}, which will pin the $C_n$ axis along either $\hat{x}$ or $\hat{y}$ direction. One can then determine $n$ by measuring the spin population along $\hat{x}$ or $\hat{y}$ \cite{supple}.

(iii) Distinguishing $D_\infty$, $T$ and $O$: Our calculations show an ultralow magnetic field always pin the $C_3$ axis of $T$ symmetry, while pins the $C_4$ axis of the $O$ state. Even though our calculations did not find the ``polar" state with $D_{\infty}$ symmetry, we note that if present, it can only be pinned along either the $C_{\infty}$ or the $C_2$ axis by magnetic field, which is distinct from $T$ and $O$. So by measurement {\bf (A)} we can easily distinguish these three. Another difference between $T$ and $O$ is that $T$ phase breaks time reversal symmetry, while $O$ phase does not. Hence for $O$ phase, we have $|\varphi_{m}|^2= |\varphi_{-m}|^2$ in zero field limit, which is a property not shared by the $T$ phase \cite{supple}.

In Fig. \ref{proposal} we summarize our proposal with a flow diagram. Following these steps,  all possible (normal point group) symmetries can be determined exactly.

{\it Final Comments:}
Although our discussions are for Dy, our method of detection is applicable for all spinor condensates such as Cr \cite{Cr}, and Er \cite{Er}.  In our discussion, we have not considered dipole energy. The competition between dipole energy and spin-dependent $s$-wave interaction will certainly change the energetics.  However, dipolar energy  is known to depend strongly on the geometry of the trap, and is known to have less effects in spherical potentials. Moreover,  since all the non-abelian states are nonmagnetic, dipolar interaction may not have strong effects on them.   In settings where dipolar interaction becomes dominant, the system tends to develop non-uniform spin textures as shown in the recent  experiments on $^{87}Rb$ \cite{Kurn}. Even in such cases, the underlying zero field states such as those discussed here still play an important role in determining the global equilibrium spin textures\cite{ho2}.

{\it Acknowledgements.} This work is supported by Tsinghua University Initiative Scientific Research Program. HZ is supported by NSFC under Grant No. 11004118 and No. 11174176, and NKBRSFC under Grant No. 2011CB921500. TLH is supported by NSF Grant DMR-0907366, and by DARPA under the Army Research Office Grant Nos. W911NF-07-1-0464, W911NF0710576.

\begin{widetext}

\section{Supplementary Material}

\subsection{Magnetic Moment of Dysprosium}
Dysprosium (Dy) atom has $10$ $f$ shell electrons, which give the atom electron spin $S=2$ and electron orbital angular momentum $L=6$ according to the first and second Hund's rule. Due to spin orbit coupling, the total spin $\mathbf{F}=\mathbf{L}+\mathbf{S}$ of Dy atom becomes $F=8$ (the third Hund's rule). The magnetic moment operator of the atom is then $\pmb{\mu}=\mu_b(\mathbf{L}+2\mathbf{S})$, where $\mu_b=5.79\times 10^{-5}eV\cdot T^{-1}$ is Bohr magneton. Typically the spin orbit coupling energy of an atom is around $10^{-2}eV$, while Zeeman energy and interaction energy per atom in experiments are usually lower than $10^{-8}eV$. Thus Dy atoms will nearly perfectly stay in the coupled $F=8$ representation, and we can use the projection of $\pmb{\mu}$ into the $F=8$ subspace as the magnetic moment.

Both $\mathbf{F}$ and $\pmb{\mu}$ are $SU(2)$ irreducible tensor operators of order $1$ (vector operators). Wigner-Eckart theorem says an order $k$ $SU(2)$ irreducible tensor operator $T_q^k$ behaves as
\begin{equation}
\langle F',m'|V^k_q|F,m\rangle=\frac{\langle F'm'kq|Fm\rangle}{\sqrt{2F+1}}\langle F'||V^k||F\rangle\ \label{Wigner}
\end{equation}
where $\langle F^\prime m^\prime kq|Fm\rangle$ is the Clebsch-Gordan coefficient, and $\langle F^\prime||V^k||F\rangle$ is a factor independent of $m^\prime,m,q$. For $F^\prime=F$, Eq. (\ref{Wigner}) then tells us all the tensor operators of spin $k$ are proportional when projected into the spin $F$ subspace. As a special case of $k=1$, the projected magnetic moment $\pmb{\mu}$ is proportional to spin $\mathbf{F}$, with a factor $\mu$ to be determined below.

By a dot product with $\mathbf{F}$ on both sides of $\pmb{\mu}=\mu\mathbf{F}$, we arrive at the following equation
\begin{equation}
\mu_b(\mathbf{L}+\mathbf{S})\cdot(\mathbf{L}+2\mathbf{S})= \mu_b\Big[\frac{3}{2}\mathbf{F}^2-\frac{1}{2}(\mathbf{L}^2-\mathbf{S}^2)\Big] =\mu\mathbf{F}^2
\end{equation}
By noticing $\mathbf{L}^2=L(L+1)$, $\mathbf{S}^2=S(S+1)$ and $\mathbf{F}^2=F(F+1)$, we can derive out
\begin{equation}
\mu=\mu_b\Big(\frac{3}{2}-\frac{L(L+1)-S(S+1)}{2F(F+1)}\Big)
\end{equation}
For Dy atom with $S=2$, $L=6$ and $F=8$, we have $\mu=5\mu_b/4$.

\subsection{Interaction Parameters and Ground State}
\begin{table}[b]
\caption{\label{g2j} Interaction parameters $g_{2j}-g_0$ which lead to ground states shown in Fig. 1.}
\begin{ruledtabular}
\begin{tabular}{c|c|cccccccc}
Phase&Isotropy group& $g_{2}-g_0$ & $g_{4}-g_0$ & $g_{6}-g_0$ & $g_{8}-g_0$ & $g_{10}-g_0$ & $g_{12}-g_0$ & $g_{14}-g_0$ & $g_{16}-g_0$ \\
\colrule
Fig. 1(a) & $C_{5v}$ &1&0&0&-1&-2&0&-2&1\\
Fig. 1(b) & $C_{\infty}$ &1&0&0&1&4&0&-5&1\\
Fig. 1(c) & $D_{8d}$ &-1&0&0&0&0&0&0&1\\
Fig. 1(d) & $D_{4d}$ &1&0&0&1&4&0&0&1\\
Fig. 1(e) & $D_{14h}$ &2&2.5&0&1.5&5&7&2.5&4\\
Fig. 1(f) & $T_d$ &0.1&0&0&-1&-1&0&0&1\\
Fig. 1(g) & $T_d$ &0.1&0&0&-1.27&-1&0&0&1\\
Fig. 1(h) & $T$ &0.1&0&0&-8&-1&0&0&1\\
Fig. 1(i) & $O_h$ &1&1&1&1&-5&-5&1&1\\
\end{tabular}
\end{ruledtabular}
\end{table}

By using the relationship $\sum_{j=0,2,\cdots}^{2F}\sum_{m=-j}^{j}|\mathcal{A}_{jm}|^2=(\varphi^\dag\varphi)^2$, the mean-field interaction energy can be rewritten as
\begin{equation}
\mathcal{E}_{int}=g_0n^2_0(\varphi^\dag\varphi)^2 +\sum\limits_{j=0,2,\cdots}^{2F}(g_{j}-g_0)n^2_0\sum\limits_{m}|\mathcal{A}_{jm}|^2 \label{Eint}
\end{equation}
where $\mathcal{A}_{jm}=\sum_{m_1}\langle j,m|F,m_1,F,m-m_1\rangle\varphi_{m_1}\varphi_{m-m_1}$. Only the second part of Eq. (\ref{Eint}) is spin dependent and useful for the determination of the ground state, which consists of $8$ parameters $g_{j}-g_0$ for $F=8$. As examples, the interaction parameters which lead to ground states shown in Fig. 1 of our paper are summarized in Table \ref{g2j} (without external magnetic field). These parameters \{$g_{j}-g_0$\} are chosen to be located at the interior of the corresponding phases' space, and therefore some variation of these parameters will not change the symmetry of the states. One should also note that a simultaneous multiplication on $\{g_{j}-g_0\}$ by a factor $\lambda$ (i.e. $\{g_{j}-g_0\}\rightarrow\{\lambda(g_{j}-g_0)\}$) do not change the ground state.


\subsection{Spin Population in a Magnetic Field}

When an ultralow magnetic field $\mathbf{B}=B\hat{z}$ is turned on, the symmetry of Hamiltonian is reduced to $SO(2)\times U(1)$, and the ground state will generally have one of its rotational axis fixed in $\hat{z}$ direction. For $C_n$ symmetry which are ferromagnetic phases, the spin $\langle \mathbf{F}\rangle$ points along $C_n$ axis. Thus the first order perturbation energy $-\mu\mathbf{B}\cdot\langle\mathbf{F}\rangle$ is non-zero, and the $C_n$ axis will be fixed along $\hat{z}$ direction. For non-magnetic phases with $D_n, T$ or $O$ symmetry, it is the second order perturbation that determines which axis is fixed toward $\hat{z}$. In Table \ref{population} the calculated spin populations (under $\hat{z}$ basis) of several phases in a ultralow magnetic field are shown, from which it is easy to figure out the fixed axis according to Measurement {\bf(A)}. In the calculations the Zeeman energy added is around $1\%\sim10\%$ of the spin dependent interaction energy, and we note this Zeeman energy only changes the magnitude of non-zero spin components, while keeps vanishing components (in the zero field limit) zero. For different states with $D_n$ symmetry, it is either the $C_n$ or the $C_2$ axis that is fixed along $\hat{z}$ direction. Numerical computations show for $D_{nd}$ ($D_{nv}$) symmetry, $C_2$ ($C_n$) axis is more likely fixed. For $T$ and $O$ symmetry, $C_3$ axis and $C_4$ axis are fixed along $\hat{z}$, respectively.


\begin{table}[h]
\caption{\label{population} Calculated population $\rho_m=\varphi_m^\dag\varphi_m$ measured along $+\hat{z}$ direction weak magnetic field. In the calculations we are adding a Zeeman energy around $1\sim10$ percents of the spin dependent interaction energy.}
\begin{ruledtabular}
\begin{tabular}{cccccccc}
Phase &Fig. 1(a)&Fig. 1(b)&Fig. 1(c)&Fig. 1(d)&Fig. 1(e)&Fig. 1(f)&Fig. 1(i)\\
\colrule
Isotropy group &$C_{5v}$&$C_{\infty v}$&$D_{8d}$&$D_{4d}$&$D_{14h}$&$T_d$&$O$\\
\colrule
Axis fixed &$C_5$&$C_\infty$&$C_2$&$C_2$&$C_{14}$&$C_3$&$C_{4}$\\
\colrule
$\rho_8\ $    &0&0&0.1191&0.1963&0&0&0.2210\\
$\rho_7\ $    &0.4922&1.0000&0&0&0.5590&0.0003&0\\
$\rho_6\ $    &0&0&0.0649&0.4341&0&0&0\\
$\rho_5\ $    &0&0&0&0&0&0&0\\
$\rho_4\ $    &0&0&0.0780&0.0170&0&0.4953&0.0717\\
$\rho_3\ $    &0&0&0&0&0&0&0\\
$\rho_2\ $    &0.4629&0&0.1559&0.2254&0&0&0\\
$\rho_1\ $    &0&0&0&0&0&0.1736&0\\
$\rho_0\ $    &0&0&0.2110&0.0583&0&0&0.5179\\
$\rho_{-1}$ &0&0&0&0&0&0&0\\
$\rho_{-2}$ &0&0&0.1451&0.0444&0&0.0604&0\\
$\rho_{-3}$ &0.0424&0&0&0&0&0&0\\
$\rho_{-4}$ &0&0&0.0693&0.0005&0&0&0.0646\\
$\rho_{-5}$ &0&0&0&0&0&0.0426&0\\
$\rho_{-6}$ &0&0&0.0567&0.0192&0&0&0\\
$\rho_{-7}$ &0&0&0&0&0.4410&0&0\\
$\rho_{-8}$ &0.0025&0&0.1000&0.0049&0&0.2279&0.1247\\
\end{tabular}
\end{ruledtabular}
\end{table}

\subsection{Effect of Magnetic Field Gradient for $D_{nd}$ Symmetry}

For $D_{n}$ phase whose $C_2$ axis is fixed along $\hat{z}$ (such phases are likely to have a $D_{nd}$ symmetry), measuring the spin populations under $\hat{z}$ basis does not give us the number $n$. To figure out $n$, we need to add an ultralow magnetic field with small gradient $\mathbf{B}=B[\hat{z}+G_0 (x\hat{x}-z\hat{z})]$. By a local unitary transformation $\varphi\rightarrow e^{-iG_0 F_y x}\varphi$, all local $\hat{z}$ axes are aligned along $\mathbf{B}$, and a kinetic energy cost \cite{Ho}
\begin{equation}
\mathcal{E}_{k}=-n_0\hbar^2\varphi^\dag e^{iG_0F_yx}\nabla^2e^{-iG_0F_yx}\varphi/(2M)=n_0\epsilon\langle F^2_y\rangle
\end{equation}
arises, where $\epsilon=\hbar^2G_0^2/(2M)$. This kinetic energy must be far smaller than Zeeman energy, for otherwise the Bose Einstein Condensate will prefer the uniform configuration that costs the Zeeman energy instead of the kinetic energy. This kinetic energy term will fix $C_n$ axis of the $D_{n}$ symmetry along either $\hat{x}$ or $\hat{y}$. First we know for $D_{n}$ phases with one of the $C_2$ axes fixed along $\hat{z}$ by external field $\mathbf{B}$, the $C_n$ axis then lies in $\hat{x}-\hat{y}$ plane. By symmetry consideration, $C_n$ must be one principal axis of the real symmetric tensor $Q_{ij}=\langle (F_iF_j+F_jF_i)\rangle/2$ ($i,j=x,y,z$). Thus $\langle F_y^2\rangle=\hat{y}\cdot \mathbf{Q}\cdot \hat{y}$ will reach the minimum when $C_n$ axis points along either $\hat{x}$ or $\hat{y}$. A Measurement of the spin populations in either $\hat{x}$ or $\hat{y}$ basis will then give us the number $n$. For the two phases $D_{8d}$ and $D_{4d}$ shown in Fig. 1(c) and (d), our calculations show that their $C_8$ and $C_4$ axes are fixed along $\hat{x}$ and $\hat{y}$, respectively.

In experiments it is hard to change the basis for spin population measurement, so an easier method is to perform a $\pi/2$ rotation of atoms about $\hat{y}$ or $\hat{x}$ axis, followed by a measurement of spin population in $\hat{z}$ basis. A $\pi/2$ rotation about $\hat{y}$ axis can be realized by adding an additional alternating magnetic field $\mathbf{B}_{a}=B_{a}\hat{y}\cos\Omega t$ where $\Omega$ is the Zeeman splitting frequency $\mu B/\hbar$. The frequency $\Omega$ will equivalently recover the degeneracy of split Zeeman levels, and the atoms will rotate an angle $\mu B_aT/\hbar$ about $\hat{y}$ axis within time $T$. $\pi/2$ rotation is realized by controlling $T=\pi\hbar/(2\mu B_a)$.

\end{widetext}


\begin{thebibliography}{99}

\bibitem{Dy}
M. Lu, N. Q. Burdick, S. H. Youn, and B. L. Lev, Phys. Rev. Lett. {\bf 107}, 190401 (2011).

\bibitem{Cr}
B. Pasquiou, E. Marchal, G. Bismut, P. Pedri, L. Vernac, O. Gorceix, and B. Laburthe-Tolra, Phys. Rev. Lett. {\bf 106}, 255303 (2011).

\bibitem{non-abelian}
R. Barnett, A. Turner, and E. Demler, Phys. Rev. Lett. {\bf 97}, 180412 (2006).

\bibitem{Mermin}
See Sec V and VI of the review article by N. D. Mermin, Rev. Mod. Phys. {\bf 51}, 591 (1979).

\bibitem{FM}
T. L. Ho, Phys. Rev. Lett. {\bf 81}, 742 (1998).

\bibitem{Machida}
T. Ohmi and K. Machida, J. Phys. Soc. Jpn., {\bf 67}, 1822 (1998)

\bibitem{supple}
See supplementary material for detail.

\bibitem{Majorana}
E. Majorana, Nuovo Cimento {\bf 9}, 43 (1932).

\bibitem{highspins}
A. Lamacraft, Phys. Rev. B {\bf 81}, 184526 (2010);
Y. Kawaguchi, and M. Ueda, Phys. Rev. A {\bf 84}, 053616 (2011);
M. Fizia, and K. Sacha, J. Phys. A: Math. Theor. {\bf 45}, 045103 (2012);
H. M{\"a}kel{\"a}, and K. -A. Suominen, Phys. Rev. Lett. {\bf 99},190408 (2007).

\bibitem{Schwinger}
R. Barnett, D. Poldolsky, and G. Refael, Phys. Rev. B {\bf 80}, 024420 (2009).

\bibitem{definition} $C_{nv}$ is $C_{n}$ plus $n$ vertical mirror planes.  $C_{nh}$ is $C_{n}$ plus a horizontal mirror plane,
$S_{2n}$ is generated by a $\pi/n$ rotation combined a horizontal mirror reflection. Operating $S_{2n}$ twice will result in a $C_n$ rotation. $D_{nh}$ is $D_n$ plus  a vertical mirror plane containing a $C_2$ axis.
$D_{nd}$ is  $D_n$ plus a vertical mirror plane between two $C_2$ axes. $T_d$ is $T$ plus a mirror plane that dissects the tetrahedron evenly. $O_h$ is $O$ plus inversion.  See J. P. Elliott, and P. G. Dawber, Symmetry in Physics, Oxiford University Press, New York, 1986.

\bibitem{guess} $C_{nh}$ and $S_{2n}$ are subgroups of $D_{nh}$ and $D_{nd}$ respectively. States with $C_{nh}$ ($S_{2n}$) symmetry but no $D_{nh}$ ($D_{nd}$) are never found in calculations.

\bibitem{reason}
If $n$ is odd and $2n>16$, then there must be $n$ points on the equator. The other $16-n$ points can only be put into the north or south pole to satisfy $C_{n}$ symmetry. However, since $16-n$ is odd. These points can not be distributed evenly at the two poles, making it impossible to have $D_n$ symmetry.

\bibitem{Er}
Laser cooling has been made for Er: A. J. Berglund, J. L. Hanssen, and J. J. McClelland, Phys. Rev. Lett. {\bf 100}, 113002 (2008); F. Ferlaino and R. Grimm, Private communication.

\bibitem{Ho}
T. L. Ho, and S. K. Yip, Phys. Rev. Lett. {\bf 84}, 4031 (2000).

\bibitem{Kurn} M. Vengalattore, S. R. Leslie, J. Guzman, and D. M. Stamper-Kurn, Phys. Rev. Lett. 100, 170403 (2008).

\bibitem{ho2} J. Zhang, and T. L. Ho, Journal of Low Temperature Physics, {\bf 161}, 325, (2010).


\end{thebibliography}
\end{document}